\documentclass[reprint,NumberedRefs]{JASA}

\usepackage{amsmath}
\usepackage{algorithm2e} 
\usepackage{gensymb}

\begin{document}
\title[JASA/Sample JASA Article]{Angle-distance decomposition based on deep learning for active sonar detection}
\author{Jichao Zhang}
\email{zhangjichao@mail.nwpu.edu.cn}
\affiliation{School of Marine Science and Technology, Northwestern Polytechnical University, Xi'an 710072, China}
\affiliation{Research and Development Institute of Northwestern Polytechnical University in Shenzhen, Shenzhen 518063, China}
\author{Xiao-Lei Zhang}
\email{xiaolei.zhang@nwpu.edu.cn}
\affiliation{School of Marine Science and Technology, Northwestern Polytechnical University, Xi'an 710072, China}
\affiliation{Research and Development Institute of Northwestern Polytechnical University in Shenzhen, Shenzhen 518063, China}
\author{Kunde Yang}
\email{ykdzym@nwpu.edu.cn}
\affiliation{Ocean Institute of Northwestern Polytechnical University, China}
%\author{Rui Duan}
%\email{duanrui@nwpu.edu.cn}
%\affiliation{Ocean Institute of Northwestern Polytechnical University, China}
%\footnotetext[5]{Xiao-Lei Zhang and Kunde Yang are corresponding authors}
%\footnote[1]{Xiao-Lei Zhang and Kunde Yang are the corresponding authors.}
%\altaffiliation{Xiao-Lei Zhang and Kunde Yang are the corresponding authors.}
%\affiliation{Xiao-Lei Zhang and Kunde Yang are the corresponding authors.}
%\preprint{Author, JASA}		
%  Used if you want want this message to appear in upper right corner of title page, for preprint only.

\date{\today}

\begin{abstract}
Underwater target detection using active sonar constitutes a critical research area in marine sciences and engineering. However, traditional signal processing methods face significant challenges in complex underwater environments due to noise, reverberation, and interference. To address these issues, this paper presents a deep learning-based active sonar target detection method that decomposes the detection process into separate angle and distance estimation tasks. Active sonar target detection employs deep learning models to predict target distance and angle, with the final target position determined by integrating these estimates. Limited underwater acoustic data hinders effective model training, but transfer learning and simulation offer practical solutions to this challenge. Experimental results verify that the method achieves effective and robust performance under challenging conditions.
\end{abstract}

\maketitle

\section{\label{sec:1} Introduction}
Underwater target detection has long been a subject of significant scientific and practical interest. Among various approaches, many studies have focused on detection methods based on passive sonar. Nevertheless, passive sonar systems inherently suffer from two fundamental limitations. First, they are incapable of detecting targets that remain completely silent. Second, passive sonar can only receive signals emitted by the target, thereby lacking access to a reference signal that could facilitate more accurate detection. In contrast, active sonar systems are well suited to address these limitations. By actively transmitting acoustic signals\citep{stewart1959theory}, active sonar not only enables the detection of silent targets but also allows the transmitted signal to be used as a reference throughout the target detection process. It has widespread applications in various fields, including ocean resource exploration \citep{6684011,li2019underwater}, environmental monitoring \citep{baldaccisignal,nagla2024sonar}, and military operations \citep{haralabus2004broadband,fallon2013relocating}. Active sonar systems operate by transmitting specific signals and receiving corresponding echoes to detect and analyze the surrounding marine environment. However, the complex underwater environment \citep{nagla2024sonar,wenz1972review}, characterized by ocean noise, reverberation, and interference from reflections of other objects, imposes significant challenges on the accuracy and robustness of traditional signal processing and classification techniques.

Traditional active sonar target recognition is based on effective feature extraction methods, such as direction-of-arrival (DOA) \citep{nie2024multi} and time-of-arrival (TOA) \citep{peters2016bayesian,he2024time} estimation. In the field of underwater DOA, among which conventional beamforming (CBF) \citep{benesty2008conventional} , minimum variance distortionless response (MVDR) \citep{cox1987robust}, and multiple signal classification (MUSIC) \citep{schmidt1986multiple} are the most commonly employed methods. CBF operates by applying phase shifts to signals received by different array elements and summing them to form a beam steered toward a specific direction. MVDR is a data-adaptive approach that minimizes the variance of the beamformed output while maintaining a distortionless response toward the target direction. MUSIC is a subspace-based super-resolution algorithm that leverages eigenvalue decomposition of the sample covariance matrix to separate the signal subspace from the noise subspace. The TOA method detects the target position according to the arrival time of the signal at different hydrophones. Although these traditional methods can be effective under specific conditions, practical applications often face significant challenges due to the complexity of the underwater environment. Background noise and reverberation may obscure target signals, further reducing the SNR. Therefore, enhancing the recognition capability of active sonar in complex environments and improving system robustness and adaptability have become critical research challenges in underwater acoustic signal processing. 

In recent years, deep learning (DL) \citep{moniruzzaman2017deep} has emerged as promising solutions to these challenges, providing new insights and methodologies for underwater target recognition. 
This method enables the efficient processing of large-scale and complex data while overcoming the limitations of traditional handcrafted feature extraction techniques. 
The study in \cite{10052708} employs a feature extractor to extract features from forward-looking sonar images and uses deep learning for localization. 
The work in \cite{wang2022mlffnet} proposes a multi-level feature fusion network that utilizes modules such as multi-scale convolution, feature extraction, and feature fusion to perform object detection based on sonar images. 
The study in \cite{najibzadeh2023active} introduces a deep convolutional neural network approach to classify active sonar images for target detection. 
Existing studies primarily focus on first transforming the received signals into images through signal processing techniques, and then using these images as inputs to the model \citep{xie2022dataset,sethuraman2024machine}. 
However, this approach may lead to the loss of certain critical information inherently contained in the original signals.
The synthesized image relies on image formation algorithms \cite{hayes2009synthetic}, such as the time-domain delay-and-sum method, which requires resampling and may reduce signal accuracy. In addition, platform motion can lead to image blurring and ghosting artifacts.
%In particular, the introduction of convolutional neural network (CNN) \citep{bouvrie2006notes} and the Transformer \citep{vaswani2017attention} framework has provided novel solutions for active sonar target recognition. Currently, CNN have been widely adopted in signal processing due to their advantages in time-frequency-angle feature extraction \citep{yang2021cnn,chakrabarty2019multi}. By leveraging convolutional layers, CNN can effectively capture sonar signal characteristics, and several studies \citep{wang2019underwater,papageorgiou2021deep} have explored their applications in sonar signal detection. Meanwhile, the Transformer architecture have also garnered significant attention. With its powerful self-attention architecture, the Transformer can model global dependencies in long-time sequences, making it particularly suited for addressing long-sequence modeling and multi-path effect challenges in underwater acoustic signal processing \citep{feng2022transformer}. 

To address this issue, we propose an active sonar target detection (ASTD) method based on Deep Neural Networks (DNN) \citep{gong2021ast}. The method decomposes the problem of active sonar target detection into two separate tasks of angle and distance estimation. First, the angle estimation task is formulated as a multi-class classification problem. A DNN model is trained using the phase characteristics of received signals from each array element, enabling the model to predict the target angle based on phase differences across different elements. Then, a Transformer-based model is employed for distance estimation, where the received signals are converted into Mel-spectrogram. The Mel-spectrogram is then divided into multiple patches, and positional encoding is introduced. The patch feature vectors, along with their positional encodings, are fed into the Transformer model, allowing it to learn the target’s characteristics effectively. The final target position is obtained by integrating the angle and distance estimates. Finally, simulation experiments are conducted to validate the effectiveness of the proposed active sonar target detection method. The contributions of this paper are outlined as follows: 

$\bullet$ We propose a novel ASTD method that decomposes the detection task into separate angle and distance estimation processes. A DNN is utilized for angle estimation based on phase differences, while a Transformer model processes Mel-spectrogram patches with positional encoding for distance estimation. The final target position is determined by fusing angle and distance estimates.

$\bullet$ The proposed method offers high flexibility by allowing the replacement of different backbone networks, making it adaptable to various architectures. Additionally, it demonstrates strong generalization capabilities, enabling its application in diverse environments and varying target numbers.

$\bullet$ Since DNN models require large amounts of training data, we address the challenge of limited and low-quality underwater acoustic data by leveraging transfer learning and simulation-based data augmentation.

The structure of the paper is as follows: In Sec. \ref{sec:2}, the active sonar signal model is formulated. In Sec. \ref{sec:3}, the ASTD method is described in detail. In Sec. \ref{sec:4}, the experimental results are analyzed and discussed. Finally, the conclusions are summarized.

\section{\label{sec:2} ACTIVE SONAR SIGNAL MODEL}
The working mode of active sonar mainly consists of two processes: transmitting signals and receiving target echoes. 
Assuming there are underwater targets from different directions represented as $s$, their direction‑of‑arrival angles(DOA) can be expressed as $(h_1, h_2, \cdots, h_s)$.

\begin{figure}[ht]
\includegraphics[width=\reprintcolumnwidth]{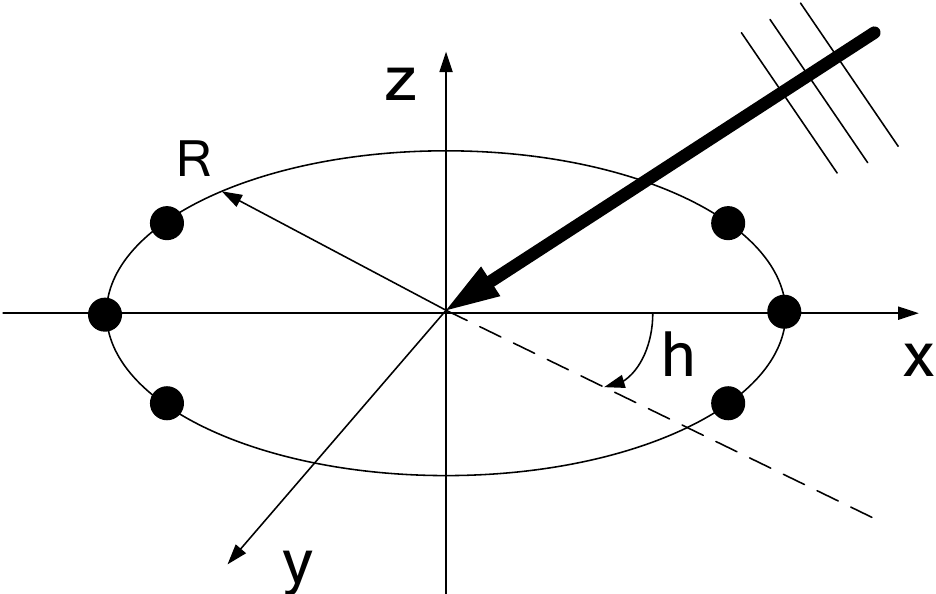}
\caption{Circular hydrophone array}
\raggedright
\label{fig.sonar}
\end{figure}

As Fig.\ref{fig.sonar} shows, these target signals are incident on a circular hydrophone array (CHA) containing $N$ array elements, and the array elements are evenly distributed along a circular ring with a radius $r$.

Receive signals $r_{j}(t)$  at times $t$ can be represented as:
\begin{equation}
r_{j} ( t ) = \sum _ { i = 0 } ^ { M } A _ { i } x ( t - \tau _ { i } ) + z ( t ),
\label{eq1}
\end{equation}
where $j \in N$, $x(t-\tau_{0})$corresponds to direct signal. When $i>0$, $x(t-\tau_{i})$ represents multi-path signals. $M$ is the number of eigenrays through the receiving array elements. $A_i$ and $\tau_i$ represents the sound pressure loss and delay relative to the transmitted signal of the $i$-th path to the receiving point, respectively. $z(t)$ represents the noise.

%After the short-time Fourier transform(STFT), the array element received signal can be obtained from Eq.~(\ref{eq1}).
%\begin{equation}
%R_{j} = A _ { 0 } X _ { 0 } + \sum _ { i = 1 } ^ { M } A _ { i } X _ { i } + Z
%\label{2}
%\end{equation}
%
%$R_{j}$ will be used as the input for the model later.

\section{\label{sec:3} DETECTION METHOD}
\begin{figure*}[ht]
\includegraphics[width=\textwidth]{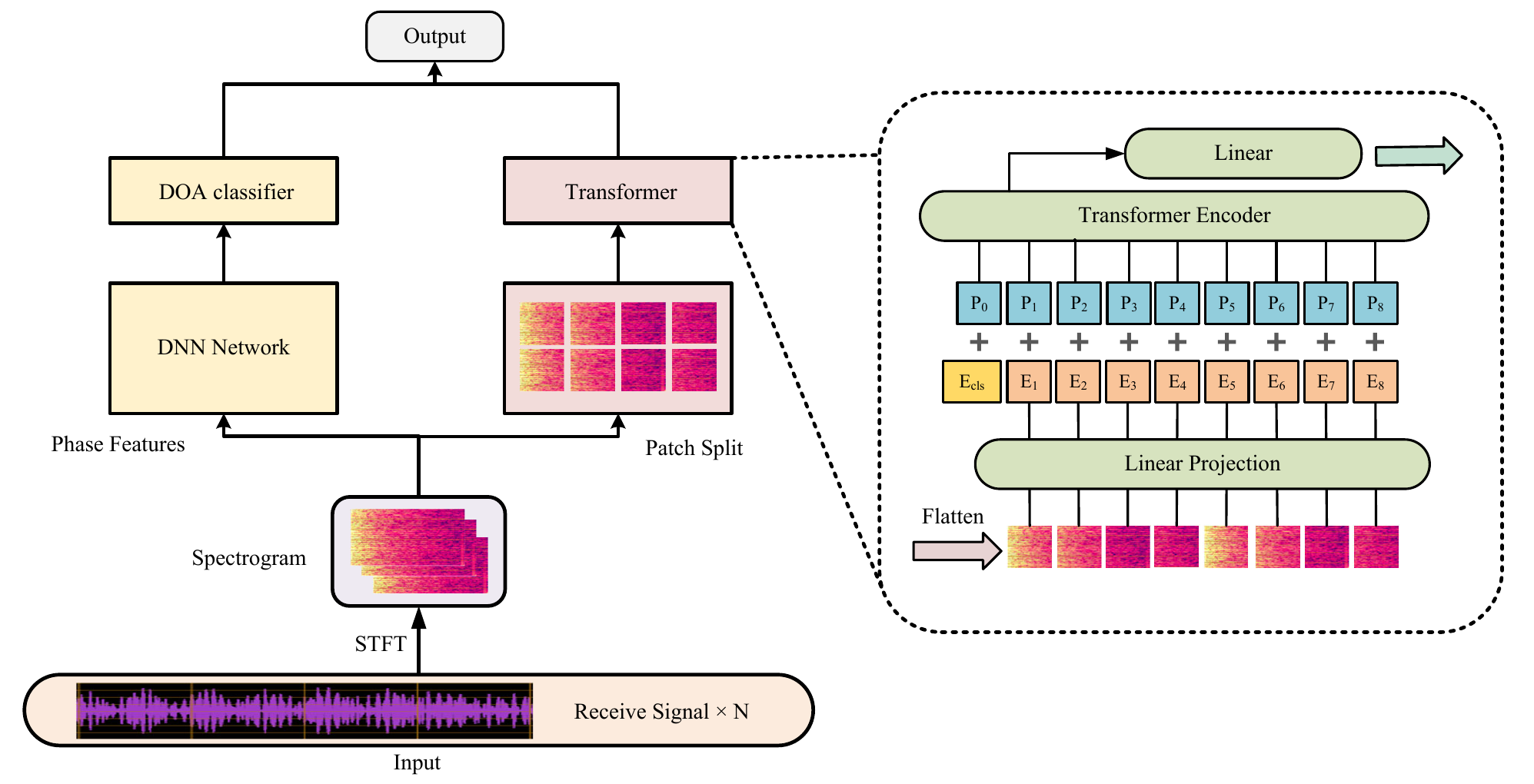}
\caption{Proposed Architecture}
\raggedright
\label{fig.block}
\end{figure*}

In this paper, Fig.\ref{fig.block} illustrate the proposed detection method architecture. This method can be divided into two parts. First, we adopt to leverage a supervised learning framework based on DNN to perform angle detection. Subsequently, a pre-training Transformer is employed to estimate the distances of multiple targets. By combining the angle and distance results, a distance-azimuth record of the targets is obtained. Please refer to the specific Algorithm\ref{alg1} for details. Then, we will introduce the functions of each module. 

\begin{algorithm}[ht]
\label{alg1}
 \caption{Activate sonar target detection task.}
    \SetKwInOut{Input}{input}
    \SetKwInOut{Output}{output}
	\Input{Receive signal: $r_j(t).$}
	\Output{Angle: $\hat{y}.$\\
             Distance: $S.$}
	 Apply short-time Fourier transform (STFT) to the received signal $r_j(t)$ \\
     \If {detection angle}
     {Extract phase components from the STFT output;\\
     Use DNN model to predict angle: $\hat{y} = u(f(r_j(t), \Theta))$.}
     \If {detection distance}
     {Split the signal into smaller segments $x_k(t)$;\\
     Use Audio Spectrogram Transformer (AST) to predict the presence of target in each segment;\\
     Convert the target segment into distance: $S = \frac{Vs \cdot T_k}{2}$.}
     Comprehensively determine the target orientation based on angle $\hat{y}$ and distance $S$.
     \vspace{0.5em}
\end{algorithm}

\subsection{Angle detection \label{subsec:3:1} }

\subsubsection{Problem transformation\label{subsubsec:3:1:1}}
The multi-target DOA estimation problem is formulated as a $y$-class multi-label classification task. First, the entire DOA range is discretized to form a set of possible DOA values, $\Theta = \{ \theta_1, \dots, \theta_y \}$. Then, a class vector $y$ is constructed, where each target corresponds to a possible DOA value in the set $\Theta$.  

In this task, we assume an independent target DOA model, meaning that the spatial locations of the targets are independent of one another. Under this assumption, the multi-label classification problem can be addressed using the binary relevance method, where the assignment of each DOA class label is treated as an independent binary classification subtask. 

Let $x(t)$ denote the input signal. Then a DNN-based DOA model $f(\cdot)$ can be expressed as:

\begin{equation}
\begin{split}
\hat{x} &= f(x(t), \Theta),\\
\hat{y} &= u(\hat{x}),
\end{split}
\label{3}
\end{equation}
where function $u(\cdot)$ maps $\hat{x}$ to a predicted distribution $\hat{y} \in [0, 1]$.

The system input is a single frame of the phase spectrum. The training objective is to make the output $\hat{y}$ as close as possible to the ground truth and minimize the loss function. During the testing phase, the input features corresponding to a single STFT phase frame into the model to obtain the $\hat{y}$, which is subsequently decoded to produce the class probabilities of the target.

\subsubsection{DNN framework\label{subsubsec:3:1:2}}
In this study, the objective is to learn features relevant to the DOA estimation task through training. Unlike other methods that directly use images as input features, we utilize phase information as the input feature representation in this research. 

The DNN framework takes the feature representation corresponding to each STFT time frame as input. After applying STFT to the received signal Eq.~(\ref{eq1}), the observed signal at each time-frequency (TF) instance is represented as a complex value. Accordingly, the observed signal is expressed as:

\begin{equation}
R_{k}(n,f) = A_{k}(n,f)e^{j\phi_{k}(n,f)}+Z(n,f),
\label{eq3}
\end{equation}
where $\phi_{k}(n,f)$ represents the phase component and $A_{k}(n,f)$ denotes the magnitude component of the STFT coefficient of the received signal at the $k$-th receiving array element for the $n$-th time frame and $f$-th frequency bin. $Z(n,f)$ represents noise.

\begin{figure}[ht]
\includegraphics[width=\reprintcolumnwidth]{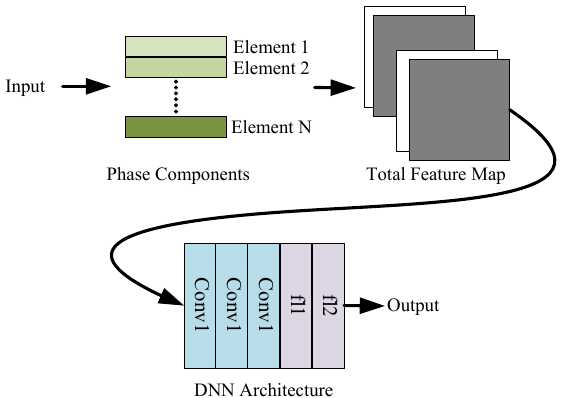}
\caption{DNN framework}
\raggedright
\label{fig.DNN}
\end{figure}

As show in Fig.\ref{fig.DNN}, we arrange the phase $\phi_{k}(n,f)$ of each receiving array element at the $n$-th time frame into a matrix of size $N\times F$, where $F$ is the total number of frequency bins, and use this matrix as input to the model. 

For details of the DNN structure, please refer to Table \ref{table:DNN}. After training, the goal is for the model to perform the DOA task based on the features extracted from the phase components.

\begin{table}[h]
\caption{DNN structure details}
\centering
\begin{ruledtabular}
\begin{tabular}{cccc}
Type&Kernel size&Input&Output\\
\hline
CONV1& 2 $\times$ 1 & 1$\times$4$\times$1501 & 4$\times$3$\times$1501\\
\hline
CONV2\&POOL& 2 $\times$ 7& 4$\times$3$\times$1501 & 16$\times$2$\times$750\\
\hline
CONV3\&POOL& 2 $\times$ 7& 16$\times$2$\times$750 & 32$\times$1$\times$375\\
\hline
FC1&-& 32$\times$1$\times$375&1$\times$1$\times$144\\
\hline
FC2&-& 1$\times$1$\times$144&1$\times$1$\times$32\\
\end{tabular}
\end{ruledtabular}
\label{table:DNN}
\end{table}

\subsection{\label{subsec:3:2} Distance detection}

\subsubsection{Problem transformation\label{subsubsec:3:2:1}}
In this task, to obtain distance information, we split the signal into smaller segments and then identify the segments that contain information related to the target. Therefore, we transform the distance detection task into a binary classification problem $p \in \{p_0, p_1\}$. Segments containing the target signal are labeled as $p_1$, while those without the target signal are labeled as $p_0$. Transformer model is used to obtain the labels. After training, the model can classify the signal segments to identify those containing the target. 

For the signal $x(t)$, it is split along the time dimension into signal segments $x_{\nu}(t)$,
\begin{equation}\label{4}
x(t) \rightarrow x_{\nu}(t).
\end{equation}

After obtaining the target information, we need to convert the time segment into distance information. It is assumed that the active sonar begins receiving signals immediately after transmitting. The distance $S$ can be obtained from:

\begin{equation}\label{5}
\begin{matrix}
S = V_{s}T_{\nu}/2, \\
T_{\nu}=cL_{\nu}(1-h),
\end{matrix}
\end{equation}
where $V_{s}$ is the sound speed in water, and $c$, $L_{\nu}$, and $h$ represent the segment length, the position of the segment in the sequence $\nu$, and the overlap rate, respectively.

\subsubsection{Transformer framework\label{subsubsec:3:2:2}}

Over the past years, transformer models has shown excellent performance in various tasks.
Transformers dispense with recurrence and convolutions in favor of self--attention, which allows for global context aggregation at every layer. 

To obtain the target location, we convert the signal segment $x_k$ into a sequence of 128-dimensional log Mel filterbank features. 
Subsequently, the spectrogram is split into a series of 16 $\times$ 16 patches, with an overlap of 6 in both the time and frequency dimensions. 
Each patch is then flattened into a 1D embedding of size 768. This linear projection layer is referred to as the patch embedding layer. 
To address the Transformer architecture's lack of positional information, we introduce a trainable positional embedding that enables the model to capture the spatial structure of the 2D audio spectrogram.

We prepend a learnable classification token [CLS] at the start of the sequence, which originally introduced in BERT \cite{devlin2019bert} for sequence‑level representation. Due to the characteristics of the attention mechanism, [CLS] tokens can aggregate global features.
We employ the original Transformer encoder architecture \cite{vaswani2017attention} without modifications, which facilitates the application of transfer learning to this task.

\subsubsection{Transfer learning\label{subsubsec:3:2:3}}

Compared to the CNN network, one disadvantage of the Transformer architecture is that it requires a large amount of data to achieve optimal performance. However, in the field of underwater acoustics, there is a challenge of data scarcity, mainly due to high costs, long acquisition times, and data security concerns. Additionally, there is currently a lack of open-source active sonar datasets. To achieve better detection results, we employ a transfer learning approach. The transfer learning method involves first pre-training the model on a large air acoustics domain dataset \citep{gemmeke2017audio}, then fine-tuning it on an active sonar dataset for underwater target detection. Although the speed of sound propagation in water is different from that in air, the propagation mechanisms are similar. Transfer learning enables features extracted from airborne acoustic data to assist in underwater acoustic target detection.

To achieve better performance through transfer learning, we selected a model pre-trained on a similar task using a large dataset. The pre-trained parameters were then loaded and the entire model was fine-tuned on the active sonar dataset.

\section{EXPERIMENT\label{sec:4}}

In this section, the effectiveness of the proposed method is evaluated using simulated data, accompanied by a description of the simulation data generation process. The performance of the proposed approach is compared with conventional methods under different signal-to-noise ratios (SNR), and the impact of array element configurations on detection results is analyzed. Additionally, the effects of the marine environment and transmit waveforms on model performance are examined.

\subsection{\label{subsec:4:1} Experiment settings}

\subsubsection{Dataset\label{subsubsec:4:1:1}}

\begin{figure}[h]
\includegraphics[width=0.5\textwidth]{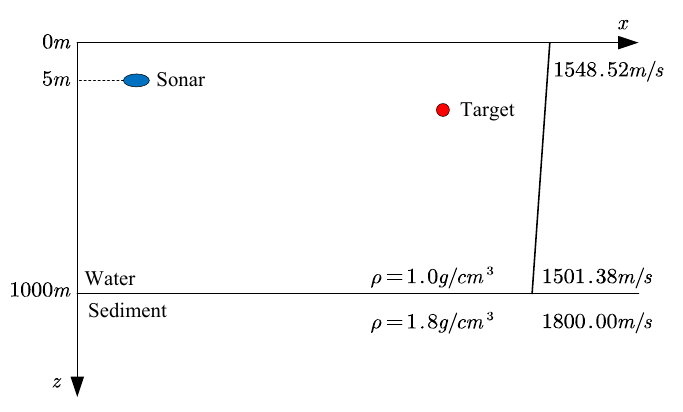}
\caption{Simulated underwater acoustic environment}
\raggedright
\label{fig.env}
\end{figure}

The simulation environment is illustrated in Fig.\ref{fig.env}, where the circular hydrophone array (CHA) is positioned at a depth of $5m$ underwater. The total water depth is $1000m$, underlain by a sediment layer. In the case of downward-refraction, the sound speed decreases from $1548.52 m/s$ at the surface to $1501.38 m/s$ at the bottom. The sound speed in the sediment layer is $1800 m/s$. The density of water is $1.0 g/cm^3$, while the density of the sediment layer is $1.8 g/cm^3$.

The CHA consists of $N=4$ hydrophones with radius $r=2.5m$, and transmits a hyperbolic frequency-modulated (HFM) signal with a starting frequency of 2000 Hz and an ending frequency of 3000 Hz, at a sample rate of 6000 Hz. The target distance from the sonar ranges from 1 to 5 km.

A total of 20,000 received signals were synthesized, each with a duration of 15 seconds and consisting of data from four array elements. For the angle estimation, the received signals from all four array elements are simultaneously fed into the DNN. For the distance estimation task, the signals are segmented into 1-second segments, and the segments containing target echoes are extracted.

\subsubsection{Parameter setting\label{subsubsec:4:1:2}}

During the distance estimation training phase, the dataset was split into training, validation, and test subsets using an 8:1:1 ratio. The model was trained for 20 epochs with an initial learning rate of $1 \times 10^{-4}$ using the Adam optimizer. A step-base learning rate scheduler with a step size of 1 epoch and a decay factor of 0.85 was employed to gradually reduce the learning rate during training.

In the angle training stage, the dataset was partitioned using the same ratio. The model was trained for 50 epochs with an initial learning rate of $1 \times 10^{-3}$, also utilizing the Adam optimizer. A ReduceLROnPlateau scheduler was applied with a reduction factor of 0.1, a patience of 3 epochs, and a minimum learning rate of $1 \times 10^{-4}$. Early stopping was implemented with a patience of 6 epochs, halting training if no performance improvement was observed.

All training and evaluation processes were conducted on a server equipped with an Intel(R) Xeon(R) Platinum 8160 CPU and four NVIDIA TITAN V GPUs.

\subsubsection{Evaluation metrics\label{subsubsec:4:1:3}}

In this paper, we employ mean absolute error (MAE), root mean square error (RMSE), Area Under Curve (AUC) and recall as metrics to evaluate model performance. MAE represents the mean absolute difference between predictions and true values. RMSE assesses prediction quality by computing the root of mean squared deviations, defined as:

\begin{equation}
\mbox{MAE} = E[\frac{1}{Y} \sum_{y = 1}^{Y}|\hat{\psi}_{y}-\psi_{y}|],
\label{eq6}
\end{equation}

\begin{equation}
\mbox{RMSE} = \sqrt{E[\frac{1}{Y} \sum_{y = 1}^{Y}(\hat{\psi}_{y}-\psi_{y})^{2}]},
\label{eq7}
\end{equation}
where $\hat{\psi}_{y}$ denotes the predicted value of the ${y}$th sample, the $\psi_{y}$ represent the truth value of the ${y}$th sample. $E[\cdot]$ represent the mean value of multiple samples. 

AUC represents the area under the Receiver Operating Characteristic (ROC) curve and measures the classifier’s performance across different threshold values. It can be expressed as

\begin{equation}
\mbox{AUC} = \frac{\sum(p_{i}, n_{j})}{\mbox{P}\times\mbox{N}},
\label{eq8}
\end{equation}
where P is the number of positive samples and N is the number of negative samples. $p_i$ is the predicted score for a positive sample, representing the probability of predicting a positive sample as positive. $n_j$ is the predicted score for a negative sample, representing the probability of predicting a negative sample as positive.

Recall measures the proportion of actual positive samples correctly identified by the model, and is defined as:

\begin{equation}
\mbox{Recall} = \frac{\mbox{TP}}{\mbox{TP}+\mbox{FN}},
\label{eq9}
\end{equation}
where TP is the number of positive class samples correctly predicted by the model, FN is the number of positive class samples incorrectly predicted as negative by the model.

\subsection{\label{subsec:4:2} Results}

\subsubsection{Main results\label{subsubsec:4:2:1}}

In this section, the performance of active sonar target detection (ASTD) is evaluated on both distance and angle estimation tasks, followed by the presentation of the final distance–azimuth map.

\begin{table}[ht]
    \caption{Distance task performance of different models at different SNR levels}
    \label{tab2}
    \centering
    \begin{ruledtabular}
        \begin{tabular}{lccc}
            Models & SNR(dB) & AUC(\%) & Recall(\%) \\
            \hline
            ASTD(ours) & 10  & 98.19  & 89.50 \\
                & 0   & 97.24  & 87.55 \\
                & -10 & 94.06 & 83.70 \\
            TDNN & 10  & 89.61 & 77.22 \\
                & 0   & 79.81 & 65.34 \\
                & -10 & 67.29 & 53.13 \\
            ResNet18 & 10  & 97.96 & 88.40 \\
                     & 0   & 95.00     & 84.34 \\
                     & -10 & 90.72 & 79.45 \\
            ResNet50 & 10  & 97.92 & 89.40 \\
                     & 0   & 95.20   & 85.32 \\
                     & -10 & 91.58  & 81.03 \\
        \end{tabular}
    \end{ruledtabular}
\end{table}

In Table \ref{tab2}, we compare the performance of ASTD with other methods, such as TDNN \citep{li2018underwater}, ResNet18, and ResNet50 \citep{he2016deep}, for the distance detection task under different SNR conditions. At an SNR of 10 dB, ASTD achieves superior performance, with an AUC of $98.19\%$ and a recall of $89.50\%$, indicating that the model effectively captures most positive samples under high signal quality conditions. In contrast, TDNN achieves an AUC of $89.61\%$ and a recall of $77.22\%$ under the same SNR, which is significantly lower than ASTD. While both ResNet18 and ResNet50 achieved high AUCs ($97.96\%$ and $97.71\%$) and Recalls ($88.4\%$ and $89.4\%$), their performance remained slightly inferior to that of ASTD. At an SNR of 0 dB, the performance metrics of TDNN and the ResNet models decline considerably. Even at an SNR of -10 dB, ASTD maintains a high AUC of $94.06\%$ and a recall of $83.70\%$, whereas the second-best model, ResNet50, achieves an AUC of $91.58\%$ and a recall of $81.03\%$ under the same condition. This demonstrates the superiority of ASTD in low-SNR environments. These results demonstrate the strong robustness of ASTD, indicating its effectiveness in capturing most samples.

\begin{table}[ht]
    \caption{Angle task performance Comparison of Different Methods at Various SNR Levels}
    \centering
    \label{tab3}
    \begin{ruledtabular}
        \begin{tabular}{lccc}
            Method & SNR(dB) & MAE(\degree) & RMSE(\degree) \\
            \hline
            ASTD (ours) & 10  & 0.95  & 2.45 \\
                        & 0   & 1.52  & 3.61 \\
                        & -10 & 1.93  & 4.18 \\
            CBF        & 10  & 12.66 & 30.06 \\
                        & 0   & 15.22 & 34.74 \\
                        & -10 & 25.55 & 46.33 \\
            MVDR       & 10  & 3.13  & 13.34 \\
                        & 0   & 9.99 & 27.90 \\
                        & -10 & 20.79 & 32.11 \\
            MUSIC      & 10  & 3.36  & 12.36 \\
                        & 0   & 11.77 & 26.75 \\
                        & -10 & 21.36 & 30.26 \\
        \end{tabular}
    \end{ruledtabular}
\end{table}

In Table \ref{tab3}, we compare the performance of ASTD with other methods, such as CBF, MVDR, and MUSIC, in the angle detection task. Under an SNR condition of -10, ASTD achieves an MAE of $1.93\degree$ and an RMSE of $4.178\degree$. It can be observed that our proposed method significantly improves both MAE and RMSE compared to traditional methods. Furthermore, traditional methods exhibit large performance variations across different SNR conditions, whereas ASTD demonstrates more stable performance across varying noise levels.

\begin{figure}[ht]
\includegraphics[width=0.5\textwidth]{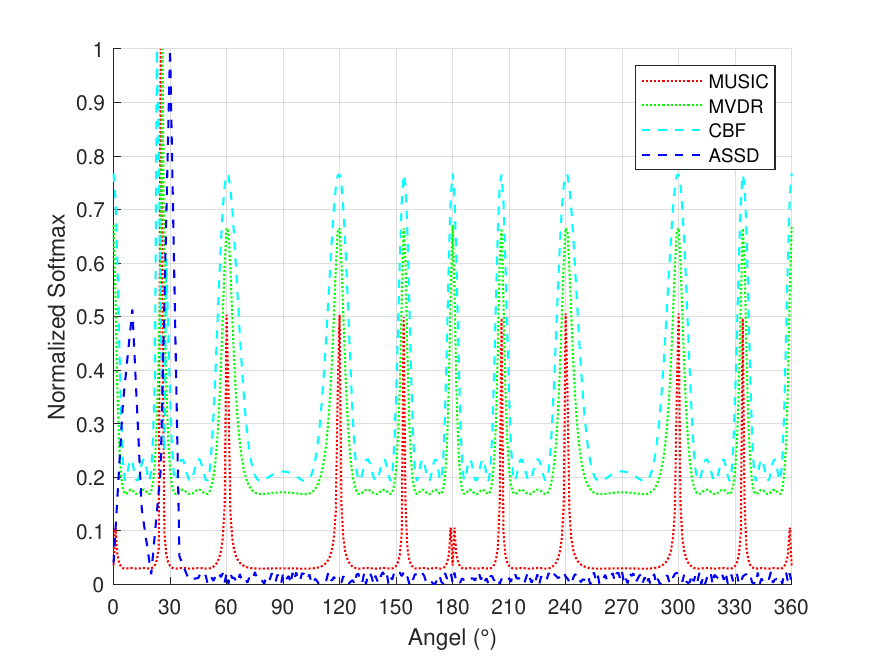}
\caption{Full bearing view normalized pseudo-spectrum}
\raggedright
\label{fig.angle}
\end{figure}

A sound source sample was generated at $30^{\circ}$ with a SNR of 10 dB. As show in Fig. \ref{fig.angle}, due to the symmetric structure of sonar arrays and the influence of reflections and multi-path effects, traditional DOA methods exhibit mirror peaks. However, based on the deep learning methods, not being affected by physical structures and similar factors, can effectively eliminate the issue of mirror peaks.

\begin{figure}[ht]
\includegraphics[width=0.5\textwidth]{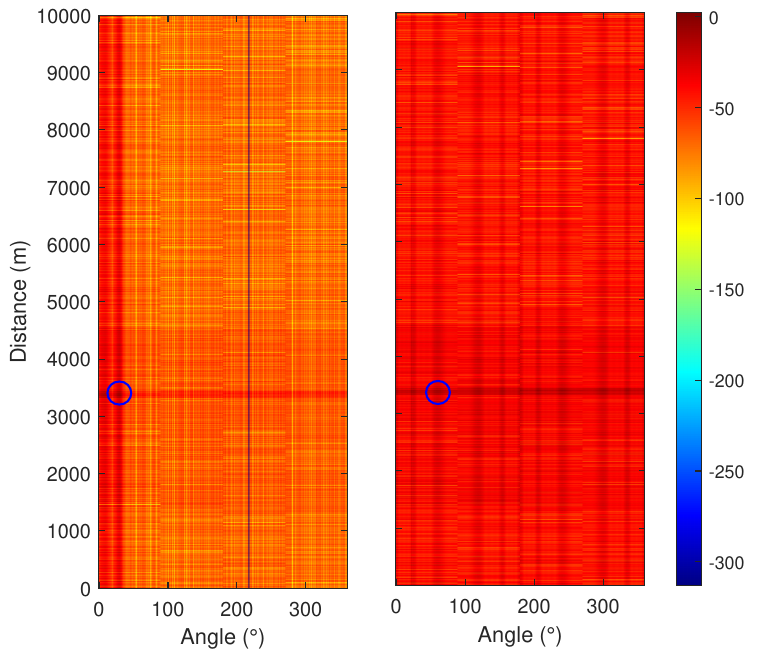}
\caption{Distance-azimuth graph. (a) Distance-azimuth graph processed by matched filtering-MVDR method. (b) Distance-azimuth graph processed by ASTD(ours) method.}
\raggedright
\label{fig.RA}
\end{figure}

We compared the performance of ASTD method and matched filtering-MVDR method. As shown in Fig.\ref{fig.RA} (a), when using the matched filtering-MVDR method, the distinction between the target signal and other interference is relatively low, making it difficult to accurately identify the target. Additionally, there is a significant deviation between the predicted target position and the actual target position, indicating that this method exhibits poor robustness in complex environments. In contrast, as shown in Fig.\ref{fig.RA} (b), the ASTD method effectively suppresses most interference signals, making the target highlights more distinguishable in the results. By leveraging deep learning, ASTD extracts more discriminative features, improving target detection capability and enhancing the separability between the target and interference. This not only increases the accuracy of target localization but also reduces the false detection rate, making the method more suitable and robust in complex acoustic environments.

\subsubsection{Effect of marine environment\label{subsubsec:4:2:2}}

\begin{table}[ht]
    \caption{Environmental parameters for different environments}
    \label{tab:environment_parameters}
    \centering
    \begin{ruledtabular}
        \begin{tabular}{lccc}
            Parameter & Env 1 & Env 2 & Env 3 \\
            \hline
            Water depth(m)  & 1000 & 100 & 2000 \\
            Sound speed profiles(m/s)  & 1548--1501 & 1500--1510 & 1500--1550 \\
            Sediment density(g/$\text{cm}^3$)  & 1.8 & 1.5 & 1.4 \\
            Sediment speed(m/s)  & 1800 & 2000 & 1450 \\
            Source depth(m)  & 100 & 10 & 100 \\
        \end{tabular}
    \end{ruledtabular}
\end{table}

\begin{table}[ht]
    \caption{MAE and RMSE for different environments}
    \label{tab:mae_rmse}
    \centering
    \begin{ruledtabular}
        \begin{tabular}{lccc}
            Environment & MAE(\degree) & RMSE(\degree) & AUC(\%) \\
            \hline
            Env 1 & 1.52 & 3.61 & 97.24 \\
            Env 2 & 1.93 & 3.66 & 96.31 \\
            Env 3 & 1.57 & 3.33 & 97.33 \\
        \end{tabular}
    \end{ruledtabular}
\end{table}

Table \ref{tab:environment_parameters} and Table \ref{tab:mae_rmse} present the environmental settings and performance under three different conditions. All other conditions were the same as in the main experiment. In Env 1, the water depth is 1000m. This results in MAE of $1.52\degree$, AUC of $97.24\%$ and RMSE of $3.61\degree$. The relatively uniform acoustic profile and deeper source position of 100m contribute to stable propagation, leading to lower errors compared to shallow water environments. In Env 2, although the water depth is only 100m. However, this setting yields the highest errors MAE of $1.93\degree$, AUC of $96.31\%$ and RMSE of $3.61\degree$. The shallow source depth of 10m and greater sensitivity to surface interactions and multi-path effects degrade model accuracy. Env 3 corresponds to the deepest setting at 2000m. The model performance is similar to Env1, suggesting improved robustness in deep-water conditions, where acoustic paths tend to be more stable. Overall, the results indicate that the model performs more reliably in deep-water environments, while shallow-water scenarios introduce increased complexity due to boundary effects, reverberation, and environmental variability.

\subsubsection{Multi-target detection\label{subsubsec:4:2:2}}

\begin{table}[ht]
    \caption{Performance under different source numbers and encoding methods}
    \label{tab:source_encoding_performance}
    \centering
    \begin{ruledtabular}
        \begin{tabular}{lccccc}
            Source No. & Encoding & MAE(\degree) & RMSE(\degree) & AUC(\%)\\
            \hline
            1 & softmax & 1.52 & 3.61 & 97.24 \\
            1 & one-hot & 2.16 & 3.68 & 97.12 \\
            2 & one-hot & 5.25 & 8.36 & 95.53 \\
            3 & one-hot & 7.42 & 15.35 & 94.82 \\
        \end{tabular}
    \end{ruledtabular}
\end{table}

Table \ref{tab:source_encoding_performance} illustrates the impact of different source numbers and encoding strategies on model performance. When the number of sources is fixed to one, the choice of encoding method has a noticeable effect on model accuracy. The softmax encoding \citep{feng2025eliminating} yields the lowest MAE of $1.52(\degree)$ and RMSE of $3.61(\degree)$, along with a slightly higher AUC of $97.24\%$ compared to one-hot encoding. This suggests that softmax encoding provides a more compact and probabilistic representation, which may enhance the model’s ability. As the number of sources increases while using one-hot encoding, both MAE and RMSE rise significantly. With two sources, the MAE increases to 5.26 and RMSE to 8.36. When three sources are present, the MAE reaches 7.42 and RMSE rises sharply to 15.35. Concurrently, AUC degrades gradually from 97.12\% to 94.82\%. These results indicate that the model performance degrades with increasing source complexity, particularly when using discrete one-hot encoding. The increase in errors attributed to difficulty in feature separation during training. Furthermore, the decline in AUC implies reduced classification confidence as more sources are introduced. Overall, the findings suggest that softmax encoding is preferable when the number of sources is limited.

\subsubsection{Transmit waveform\label{subsubsec:4:2:2}}

\begin{table}[ht]
\caption{Performance metrics under different transmit waveforms}
\label{tab:four}
\centering
\begin{ruledtabular}
    \begin{tabular}{lccc}
    Transmit Waveform & MAE(\degree) & RMSE(\degree) & AUC(\%) \\
    \hline
    CW  & 6.86 & 8.11 & 95.33 \\
    LFM & 2.53 & 6.94 & 96.78 \\
    HFM & 1.52 & 3.61 & 97.24 \\
    \end{tabular}
\end{ruledtabular}
\end{table}

Table \ref{tab:four} presents the performance of the proposed method under different transmit waveforms, including Continuous Wave (CW), Linear Frequency Modulation (LFM), and Hyperbolic Frequency Modulation (HFM). The results indicate that the choice of transmit waveform significantly impacts detection accuracy. Specifically, the CW waveform yields the highest prediction errors, with a MAE of $6.86\degree$ and a RMSE of $8.11\degree$, alongside the lowest AUC of $95.33\%$. This suggests that the CW waveform lacks sufficient diversity, making it less effective for robust target recognition. In contrast, the LFM waveform improves overall performance, MAE of $2.53\degree$ and AUC of $96.78\%$. This improvement can be attributed to the linear frequency variation of LFM signals, which enhances the richness of signal features and facilitates better model learning. Notably, the HFM waveform achieves the best results across all metrics, with an MAE of $1.52\degree$, RMSE of $3.61\degree$, and AUC of $97.24\%$. The superior performance of HFM is likely due to its nonlinear frequency characteristics, which provide improved resolution and robustness against reverberation and multi-path effects. These results demonstrate that HFM waveforms are more suitable for active sonar target detection tasks in complex underwater environments.

\section{CONCLUSION\label{sec:5}}

This work has proposed an ASTD method based on DNN and Transformer, which separately estimates angle and distance to locate underwater targets. Phase differences are used for angle prediction, while Mel-spectrograms with positional encoding are used for distance estimation. Transfer learning and simulation have been applied to address data scarcity. In experiment section, a comparative analysis has been conducted between the proposed method and other approaches, examined performance differences under varying signal-to-noise ratios, ocean environments, multiple targets, and different transmit waveforms.

\section*{ACKNOWLEDGMENTS}

\section*{AUTHOR DECLARATIONS}

\subsection*{Conflict of Interest}
The authors state that they have no conflicts to disclose.

\section*{DATA AVAILABILITY}
The data that support the findings of this study are available from the corresponding author upon reasonable request.

\bibliography{sampbib}

\end{document}